# The Angular Resolution of Space-Based Gravitational Wave Detectors

Thomas A. Moore

*Department of Physics and Astronomy, Pomona College, Claremont, CA, 91711*

Ronald W. Hellings

*Jet Propulsion Laboratory, Pasadena, CA, 91109*

(July 7, 2018)

Proposed space-based gravitational wave antennas involve satellites arrayed either in an equilateral triangle around the earth in the ecliptic plane (the ecliptic-plane option) or in an equilateral triangle orbiting the sun in such a way that the plane of the triangle is tilted at 60° relative to the ecliptic (the precessing-plane option). In this paper, we explore the angular resolution of these two classes of detectors for two kinds of sources (essentially monochromatic compact binaries and coalescing massive-black-hole binaries) using time-domain expressions for the gravitational waveform that are accurate to 4/2 PN order. Our results display an interesting effect not previously reported in the literature, and particularly underline the importance of including the higher-order PN terms in the waveform when predicting the angular resolution of ecliptic-plane detector arrays.

95.55.Ym, 04.80.Nn, 95.75.Pq, 97.60.Jd, 98.35.Jk

# I. INTRODUCTION

The era of gravitational astronomy is nearly here. In the next few years, ground-based laser gravitational wave detectors will begin to come on line and, it is hoped, space-based detectors will soon follow. Even if the ground-based detectors do not succeed in detecting the waves within the next decade, the space-based detectors will certainly be able to, so the physics goal of the detection of the new phenomenon will be achieved one way or the other, and gravitational wave detectors will then move from being physics experiments to being tools of observational astronomy.

It can be argued that the frequency band accessible to the space detectors is the more interesting one, both experimentally and observationally. It is here, at frequencies between $10^{-4}$ Hz and 1 Hz, that identified individual sources (such as AM $CVn$) exist, thus guaranteeing the detection of the phenomenon. But it is also at these frequencies that astronomical objects of the greatest observational interest are to be found, objects for which only gravitational astronomy can provide data that bears on critical astrophysical and cosmological questions. In this paper, we will investigate the ability of space-based detectors to determine the locations in the sky of two of these kinds of astronomical sources—close compact binaries in the Galaxy and coalescing massive black hole binaries in the nuclei of galaxies at high redshift.

## A. Gravitational Astronomy Science Goals

There are important questions relative to the statistics and distribution of close compact binaries in the Galaxy. Because of their low luminosity, only a few nearby white dwarf binaries have been observed, and even fewer neutron star binaries have been observed, due mainly to their narrow beams as radio pulsars. However, these endpoints of massive close binary evolution are the touchstone for the binary evolutionary models. Good average density statistics are needed in order to determine parameters of the models and to provide the basis for calculation of the branching ratios between close binary formation and other possible events such as type I supernovae. Space gravitational wave detectors are omnidirectional instruments that will provide a complete survey of close binaries. However, determining the statistics depends on measuring the spatial density, which is determined by solving for the positions of the objects that are seen gravitationally. The ability of a space detector to determine the position of the binary system it is observing is therefore an important aspect of the scientific value of the detector.

Although more speculative than the close compact binaries, there is another source that is expected to be even more interesting for physics, astronomy, and cosmology. Evidence continues to mount for the existence of massive black holes in the nuclei of most galaxies, including our own, and there is increasing evidence that most galaxies have been involved at some time in their past in a collision and merger with another galaxy. If both of these things are true, then an exciting source of gravitational waves is predicted—the coalescence of a black hole binary, formed by the collision of galaxies and the subsequent sinking of the black holes to a common center via dynamical friction. The event rate to be expected is a few per year for events out to $z = 3$. These sources will be so bright, even at this distance, that they will dominate the noise spectrum in the sensitivity window of the instruments and will be seen for months prior to the final coalescence event. In addition, if the massive black holes themselves initially grew by many small mergers of this type, then event rates of several tens of visible events per year would be expected. Observations of coalescences will provide valuable physics, since the observed waveforms represent ground truth for the computer codes presently being constructed to model the dynamics of these strong-field events. They will also provide insight into galactic merger rates, galactic and protogalactic cosmogeny, etc..

Much of the scientific value from the detection of such an event depends simply on observing the signal and determining from the waveform what the parameters of the source are, regardless of where the source is located. However, an important aspect of the science of the coalescences could also be to detect an optical counterpart of the galaxy in which the event is taking place, to see if there are observable effects produced in the galaxy by such an energetic event in its nucleus. For this goal, a position for the source would need to be determined. As we shall see, the predicted angular resolution of the space detectors will probably not be good enough to identify a single galaxy associated with the coalescence. However, it should be possible to identify the correct portion of the sky well before the final event, so that the region may be monitored and the galaxy identified by looking for some other anomoly in it. Finally, if there are many events seen per year, a knowledge of the spatial distribution of coalescences would be of value in understanding how they relate to galaxy formation statistics.

The purpose of this paper is to investigate the positional sensitivity of space gravitational wave detectors when observing signals from these two classes of sources. With signals at the expected levels for realistic sources and with noise at the levels expected for the instruments, we perform a least-squares covariance study for the parameters representing the direction to the source.

### B. Space-Based Gravitational Wave Detectors

The geometry of the detectors is shown in Figure 1. Three spacecraft[1] follow roughly circular orbits at the vertices of an equilateral triangle inscribed on the circle. The spacecraft at each of the vertices can function as the central mirror of a Michelson interferometer, while the two far spacecraft that it tracks function as the end mirrors of such an instrument. The passage of a gravitational wave changes the curvature of space between the three spacecraft and affects the phase of the laser signals passing between them, producing a signal in the interferometer. Two of the three interferometers are linearly independent of each other and provide independent information on the gravitational waveform.

The response of these detectors to a passing gravitational wave depends on the orientation of the interferometer relative to the source. We consider two different orientation scenarios. In the first, the plane of the orbit is fixed in space and the triangular geometry simply rotates in the plane at some orbital period. This is the geometry of the proposed OMEGA mission in which the spacecraft orbit the earth in a plane close to the ecliptic. In the second scenario, the plane of the orbit precesses and the triangle rotates in the precessing plane with the same period. This is the geometry that has been studied for the LISA mission, acheived by placing each spacecraft in a heliocentric orbit that is slightly eccentric and slightly inclined, so that the spacecraft all remain in a plane that is inclined 60° to the ecliptic. The dynamics of the heliocentric orbits then cause this plane to precess about the normal to the ecliptic once per year. Both orbits are sensitive to the position of the source as a result of their motion.

## II. DIRECTIONAL SENSITIVITY

### A. Extracting Directional Information

It is very difficult to determine the direction to a source that emits only a short burst of gravitational waves. Since the magnitude of the detected signal depends on both the amplitude of the wave and its direction of propagation relative to the detector arms, one would need to know the intrinsic amplitude of the wave (which would involve, for example, knowing accurately the distance to the source) to say much about its direction. Even then, the direction of the source is underdetermined, as there are many different propagation directions that yield the same signal strength. However, the orbiting-binary sources considered in this paper produce gravitational wave trains that have predictable (slowly changing) frequency and amplitude and undergo many oscillations during the course of a year of observation. Over this time span, two distinct effects modulate the detected signal in a manner that depends on a source's position in the sky.

The first of these effects is the Doppler shift due to the detector's orbital motion around the sun. The phase of this Doppler shift depends on the azimuth angle of the source in the plane of the ecliptic (the azimuthal direction of the source, that is, the direction of the projection on the ecliptic plane of a vector pointing toward the source, is the direction that the detector is moving when the source is maximally blue-shifted). The amplitude of this yearly Doppler shift depends on the polar angle of the source relative to the zenith of the ecliptic (clearly, if the source were at the zenith, there would be no Doppler shift at all, while if the source were in the plane of the ecliptic, the amplitude of the Doppler shift would be maximal). So measuring the phase and amplitude of the observed Doppler shift of the wave train allows one to determine (to some accuracy) the angular coordinates of the source.

If the detectors are deployed in the precessing-plane configuration, then a second effect comes into play. As previously discussed, the normal vector to the plane of the detector array in the precessing-plane case sweeps around the sky as the array orbits the sun. Since the detected signal depends on the direction of the wave propagation relative to the detector array, its amplitude will thus be modulated in a distinctive way as the plane precesses. Monitoring the change in signal amplitude as a function of time thus gives us additional information about the angular coordinates of the source. Since the normal to the detector plane is fixed in the ecliptic-plane case, one gets no information from this effect in that case.

The quality of the information that can be extracted from the Doppler-shift effect improves as the frequency of the source increases (having a larger number of wave cycles during a given time improves one's ability to determine the Doppler shift). The quality of the information obtained from the second effect, by contrast, is comparatively insensitive to frequency (as long as the wave still undergoes many cycles per year). We will see in Section III that our results are consistent with these general principles.

### B. Goals for This Project

The question of how well one can determine the angular coordinates of binary sources using these effects has been previously examined by Cutler [1] and Cutler and Vecchio [2,3]. Cutler's first paper discusses the an-gular resolution of precessing-plane (LISA) option only, while the papers by Cutler and Vecchio mention briefly the ecliptic-plane (OMEGA) option as well. In all of these papers, the gravitational wave is analyzed in the frequency regime, and while the wave phase is modeled to 3/2 post-Newtonian order (in the first paper) or to 4/2 PN order (in the latter two), the gravitational waveform itself is modeled only to lowest PN order (that is, it includes only a single harmonic). Also in these previous papers, the parameter space associated with the angular parameters is sampled randomly, and (in dealing with very distant events) the universe was assumed to be empty for the sake of simplicity.

---

[1]The currently proposed LISA mission presently has just such a configuration, in which each of the three spacecraft must simultaneously track in two directions. The OMEGA mission and a previous configuration of the LISA mission use two separate spacecraft at each vertex of the geometry, so that each spacecraft can rigidly orient itself to track a single direction.

In the work described here, we sought to extend and improve on this previous work by

1. Extending the approximation to 4/2-PN order for both the phase and the waveform,

2. Treating the universe as being flat ($\Omega = 1, \Lambda = 0$) instead of empty ($\Omega = \Lambda = 0$),

3. Systematically exploring the parameter space to expose patterns, and

4. Comparing the precessing-plane and the ecliptic-plane cases more fully.

We also opted to work in the time domain instead of the frequency regime and calculate all derivatives analytically instead of numerically.

Our effort to include higher-order terms in the waveform was driven by a desire to more accurately model the coalescing black-hole scenario, and it turns out (as discussed in section V below) that including these terms is especially important in the ecliptic-plane case. Although it is more complicated, we considered the flat, Friedmann universe to be a better model of the real universe than the empty universe (the distinction might be important for handling black-hole coalescence events at very large redshifts). We also felt that systematically exploring the parameter space would help us understand the behavior of these two detector configurations more physically.

## III. METHOD

### A. The Received Gravitational Wave to 4/2 Order

Blanchet, *et. al.* [4] provide a complete description of the gravitational waves emitted by inspiraling compact binaries to 4/2-PN order, assuming that the objects have zero spin, are essentially point masses, and are in a quasicircular orbit. The waveform as observed in the frame of the detector array can be expressed as follows:

$$h_{+,\times}(t) = \frac{2\tau c\eta}{5r}\varepsilon^2 \left[ H^{(0)}_{+,\times} + \varepsilon H^{(1/2)}_{+,\times} + \varepsilon^2 H^{(2/2)}_{+,\times} \right.$$
$$\left. + \varepsilon^3 H^{(3/2)}_{+,\times} + \varepsilon^4 H^{(4/2)}_{+,\times} \right] \quad (3.1)$$

where $r$ is the geometrical distance to the source, $c$ is the speed of light,

$$\tau \equiv \frac{5G(m_1 + m_2)}{c^3} \quad (3.2\text{a})$$

expresses the system's total mass in units of time,

$$\eta \equiv \frac{m_1 m_2}{(m_1 + m_2)^2} \quad (3.2\text{b})$$

is a unitless expression of the ratio of the objects' masses ($\eta = 1/4$ when the masses are equal but $\eta \to 0$ as one of the masses becomes very large compared to the other),

$$\varepsilon \equiv \left[\frac{G(m_1 + m_2)\omega}{c^3}\right]^{1/3} = \left(\frac{\tau\omega}{5}\right)^{1/3} \quad (3.2\text{c})$$

(where $\omega$ is the time-dependent angular frequency of the binary's orbit in its own frame) is a time-dependent expansion parameter that is of the same order of magnitude as the ratio $v/c$ for the system, and

$$H^{(0)}_+ = -(1 + \cos^2 i) \cos 2\phi_r \quad (3.3\text{a})$$

$$H^{(0)}_\times = -2 \cos i \, \sin 2\phi_r \quad (3.3\text{b})$$

$$H^{(1/2)}_+ = -\frac{\delta}{8} \sin i \left[(5 + \cos^2 i) \cos \phi_r \right.$$
$$\left. - 9(1 + \cos^2 i) \cos 3\phi_r\right] \quad (3.3\text{c})$$

$$H^{(1/2)}_\times = -\frac{3\delta}{4} \cos i \sin i \left[\sin \phi_r - 3 \sin 3\phi_r\right] \quad (3.3\text{d})$$

etc.

The complete expressions for the higher-order $H$s are found in Blanchet, *et. al.* [4]. In equations 3.3, $i$ is the angle of the source's orbital angular momentum vector relative to a unit vector pointing from the source to the detector ($i$ is fixed if there is no spin),

$$\delta \equiv \frac{m_1 - m_2}{m_1 + m_2} \quad (3.4)$$

is a unitless expression of the difference between the binary masses (which ranges between $-1$ and $+1$), and $\phi_r$ expresses the gravitational wave phase as a function of time. In addition to the variables $\delta, i$, and $\phi_r$, the expressions for the higher-order $H$s also depend on $\eta$. Note that the variables $\eta$ and $\delta$ are *not* independent of each other:

$$\eta = \tfrac{1}{4}(1 - \delta^2) \quad (3.5)$$

Since $\eta$ does not uniquely specify the mass ratio (for example, $m_1 = bm_2$ and $m_2 = bm_1$ yield the same $\eta$), it is better to treat $\delta$ as independent and calculate $\eta$ from it using equation 3.5.

In the frame of the *source*, the emitted gravitational wave phase $\phi_s(t_s)$ is related to the phase $\phi(t)$ of the binary orbit as follows:

$$\phi_s(t_s) = \phi(t_s) - \frac{2G(m_1 + m_2)}{c^3} \ln\left(\frac{\omega}{\omega_0}\right) \quad (3.6a)$$

where $\omega_0$ is the orbital frequency at time $t = 0$ and $t_s$ is coordinate time measured in the source frame. As discussed on page 579 in Blanchet, *et. al.* [4], the logarithmic correction term, while it is formally of 3/2 PN order, is actually of order 8/2 PN relative to the leading term $\phi(t_s)$, and thus will be very small when $\varepsilon \ll 1$. Therefore, we can safely make the approximation

$$\phi_s(t) \approx \phi(t) \quad (3.6b)$$

which significantly simplifies the calculations.

In the frame of the detector, the *received* wave is Doppler-shifted due to both the detector's orbital motion and the source's radial motion with respect to the solar system, which we assumed was entirely cosmological. The received wave frequency $\omega_r \equiv d\phi_r/dt$ depends on the binary's orbital frequency $\omega_s \equiv d\phi/dt_s$ as follows:

$$\omega_r \equiv \frac{d\phi_r}{dt} = \frac{1}{1+z}\left[1 - \frac{\Omega R}{c}\sin\Theta\sin(\Omega t - \Phi)\right]\omega_s \quad (3.7)$$

Integrating both sides of this expression yields:

$$\phi_r(t) = \frac{1}{1+z}\left[\phi(t_s) - \phi_{0s} - \frac{\Omega R}{c}\sin\Theta I_0(t)\right] + \phi_0 \quad (3.8a)$$

where

$$I_0(t) \equiv \int_0^t \omega_s \sin(\Omega t - \Phi)\, dt \quad (3.8b)$$

and where $z$ is the cosmological redshift factor, $t$ is coordinate time in the detector frame (note that $t_s = [1+z]^{-1}t$), $\phi_0$ is the phase of the received wave at time $t = [1+z]t_s = 0$, $\phi_{0s}$ is the phase of the binary's orbit at that instant, $\omega_s = \omega_s(t_s)$ is the angular frequency of the binary's orbit in its own frame at time $t_s$, $\Theta$ and $\Phi$ are the source's angular coordinates relative to the ecliptic, $\Omega$ is the angular frequency of the earth's orbit around the sun, and $R$ is the radius of the earth's orbit. (We treat both $\Omega$ and $R$ as being constant; thus $\Omega R$ is the Earth's constant orbital speed.) Equations 3.8 implicitly define $\Phi = 0$ to be the same direction as that of a displacement from the sun of the detector array's center of mass at time $t = 0$. (As mentioned before, the polar angle $\Theta$ corresponds to the zenith of the ecliptic.) Note also that since the earth's orbital speed $\Omega R/c \approx 10^{-4} \ll 1$, the Doppler shift has only been computed to first order in $\Omega R/c$.

According to Blanchet, *et. al.* [4], the orbital phase, in turn, is given by:

$$\phi(t_s) - \phi_{0s} = \frac{1}{\eta}\left[F(t_s) - F(0)\right] \quad (3.9a)$$

where

$$F(t_s) = G^5 + \eta_1 G^3 - \frac{3\pi}{4}G^2 + \eta_2 G \quad (3.9b)$$

$$G(t_s) \equiv \left[\frac{\eta}{\tau}(t_c - t_s)\right]^{1/8} \quad (3.9c)$$

$$\eta_1 \equiv \frac{3,715}{8,064} + \frac{55}{96}\eta \quad (3.9d)$$

$$\eta_2 \equiv \frac{9,275,495}{14,450,688} + \frac{284,875}{258,048}\eta + \frac{1,855}{2,048}\eta^2 \quad (3.9e)$$

where $t_c$ is the time to coalescence from $t_s = 0$, as measured in the source frame. (Specifying $t_c$ is equivalent to specifying the initial separation of the orbiting masses.) Taking the time-derivative of this function gives the orbital angular frequency $\omega_s \equiv d\phi/dt_s$ in the source frame. Note that $\omega_s$ is specified in terms of time $t_s$ in the source frame, so in the integration described in equation 3.8b. the value of the integration variable $t$ has to be transformed to $t_s$ in order to evaluate $\omega_s$ correctly.

### B. Handling Cosmological Distances

When dealing with cosmologically distant sources, it is more practical to specify the distance to the source in terms of the luminosity distance $R_L \equiv [L/4\pi F_{ob}]^{1/2}$ (where $L$ is the source's intrinsic luminosity and $F_{ob}$ is the observed flux) or the redshift factor $z$ instead of the geometrical distance $r$. In a flat, Friedmann ($\Lambda = 0$) universe, $r$, $R_L$, and $z$ are related as follows:

$$r = \frac{R_L}{1+z}, \qquad R_L H = 2\left[1 + z - \sqrt{1+z}\right] \quad (3.10)$$

where $H$ is the current value of the Hubble constant (we used $H = [14 \times 10^9 \text{ y}]^{-1}$ in our calculation). We will consider the luminosity distance $R_L$ to be the fundamental distance variable: if in some cases it is more practical to work with $z$, one can easily use equation 3.10 to change variables.

Equations 3.1 to 3.10 therefore allow us to calculate (to 4/2 PN order) both the plus and cross polarizations of the received gravitational wave (even if the source is at cosmological distances) if we know the eight quantities $\tau, t_c, R_L, \delta, i, \phi_0, \Theta$, and $\Phi$.

## C. The Detected Signal

We work in the low-frequency limit in which the phase change $\delta\phi(t)$ of the detected laser tracking signal is directly proportional to the amplitude of the wave:

$$\frac{\phi(t)}{\nu T} = h(t) = F_+(t) h_+(t) + F_\times(t) h_\times(t) \qquad (3.11)$$

where $\nu$ is the laser frequency and $T$ is the light-travel-time along the interferometer arm. The time-dependent functions $F_+(t)$ and $F_\times(t)$ are the beam-pattern functions for the interferometer pair. For both the precessing-plane case and the ecliptic plane case, the arms in a given interferometer pair make a 60° angle with respect to each other, and the beam-pattern functions for such a pair are

$$F_+ = \tfrac{\sqrt{3}}{2}\left[\tfrac{1}{2}(1+\cos^2\theta_D)\cos 2\phi_D \cos 2\psi_D \right.$$
$$\left. - \cos\theta_D \sin 2\phi_D \sin 2\psi_D\right] \qquad (3.12a)$$

$$F_\times = \tfrac{\sqrt{3}}{2}\left[\tfrac{1}{2}(1+\cos^2\theta_D)\cos 2\phi_D \sin 2\psi_D \right.$$
$$\left. + \cos\theta_D \sin 2\phi_D \cos 2\psi_D\right] \qquad (3.12b)$$

where $\theta_D$ and $\phi_D$ are the instantaneous angular coordinates of the source measured relative to the frame of the detector and $\psi_D$ specifies the orientation of the binary's principal polarization axes around the fixed line of sight. The latter variable can be understood as follows. If the orbital inclination $i$ to the line of sight is not zero or $\pi$, the quasicircular binary orbit will look elliptical to a viewer at the detector. The angle $\psi_D$ specifies the orientation of the major axis of the ellipse as viewed by this observer measured in the plane perpendicular to the line of sight and from a reference direction that is perpendicular to both the line of sight and the normal to the plane of the detector array. If $\hat{L}$, $\hat{n}$, and $\hat{z}_D$ are unit vectors parallel to the binary's conserved angular momentum, the direction of the line of sight, and the normal to the detector array, respectively, then

$$\tan\psi_D = \frac{\hat{L}\cdot[\hat{n}\times(\hat{z}_D\times\hat{n})]}{\hat{L}\cdot(\hat{z}_D\times\hat{n})} \qquad (3.13)$$

In the ecliptic-plane case, $\hat{z}_D$ is constant, so we will have $\psi_D \equiv \psi$, which defines the fixed angle of the binary orbit's major axis relative to the plane of the ecliptic. In the precessing-plane case, $\hat{z}_D$ varies with time, so we will end up having to specify $\psi_D$ as a function of $t$, $\Theta$, $\Phi$, and the fixed angle $\psi$.

In both the precessing-plane and ecliptic plane case, the equilateral arrangement of detector satellites means that the detector array possesses two independent pairs of interferometer arms, so we will have two independent signals $h_k(t)$ (where $k = 1, 2$). Because one pair of arms is rotated 60° relative to the other in the plane of the detector array, the value of $\phi_D$ will depend on the choice of interferometer pair as well, and so is denoted $\phi_{D,k}$.

For the ecliptic plane case, the zenith of the detector plane is the same as the zenith of the ecliptic, so $\theta_D = \Theta$ and, as we said, $\psi_D = \psi$. If the satellites orbit the Earth with an angular frequency of $\omega_d$, then the apparent azimuth of the source relative to the detector arm will be given by

$$\phi_{D,k} = \alpha_k(t) \quad \text{where} \quad \alpha_k(t) \equiv \Phi - \omega_d t + \alpha_{0k} \qquad (3.14)$$

Here the constant $\alpha_{0k}$ specifies the orientation of the interferometer pair at $t = 0$. While we must have $\alpha_{02} = \alpha_{01} + \tfrac{\pi}{3}$, the constant $\alpha_{01} \equiv \alpha_0$ is arbitrary. In terms of these variables, the beam-pattern functions become

$$F_{+,k} = \tfrac{\sqrt{3}}{2}\left[\tfrac{1}{2}(1+\cos^2\Theta)\cos 2\alpha_k \cos 2\psi \right.$$
$$\left. - \cos\Theta \sin 2\alpha_k \sin 2\psi\right] \qquad (3.15a)$$

$$F_{\times,k} = \tfrac{\sqrt{3}}{2}\left[\tfrac{1}{2}(1+\cos^2\Theta)\cos 2\alpha_k \sin 2\psi \right.$$
$$\left. + \cos\Theta \sin 2\alpha_k \cos 2\psi\right] \qquad (3.15b)$$

In the precessing-plane case, the orientation of the detector plane changes with time, making the expressions for the beam pattern functions more complicated. As shown in Cutler [1] (and verified by us), to get the beam-pattern functions for the precessing plane case, we must substitute into equations 3.12 the quantities

$$\cos\theta_D = \tfrac{1}{2}\cos\Theta - \tfrac{\sqrt{3}}{2}\sin\Theta\cos\beta \qquad (3.16)$$

$$\phi_{D,k} = \alpha_k(t) + \tan^{-1}\left[\frac{\sqrt{3}\cos\Theta + \sin\Theta\cos\beta}{2\sin\Theta\sin\beta}\right] \qquad (3.17)$$

where in this case

$$\alpha_k(t) \equiv \Omega t + \alpha_0 k \qquad (3.18)$$

and $\beta(t)$ specifies the angular position of the detector array's center of mass in the plane of the ecliptic. Taking account of the way that we have defined $\Phi = 0$, the quantity $\beta(t)$ is given by:

$$\beta(t) \equiv \Omega t - \Phi \qquad (3.19)$$

One can also show that $\psi_D$ in this case is given by

$$\tan\psi_D = \frac{-a\cos\psi + b\sin\psi}{a\sin\psi + b\cos\psi} \qquad (3.20)$$

where

$$a \equiv \sqrt{3}\sin\beta \qquad (3.21a)$$

$$b \equiv \sqrt{3}\cos\theta\cos\beta + \sin\theta \qquad (3.21b)$$

To summarize, we see that the detected signal from a inspiralling binary source depends on nine parameters $\tau$, $t_c$, $R_L$ (or $z$), $\eta$ (or $\delta$), $i$, $\phi_0$, $\psi$, $\Theta$, and $\Phi$, each of which we would like to determine from the observed signal. In addition, we have one parameter $\alpha_0$ that we will know independently of the signal (once the array is launched, we should know the orientation of the interferometer arms as a function of time and thus $\alpha_0$ at the given instant we define to be $t = 0$).

### D. Statistical Methods

The goal of our analysis is to estimate the accuracy with which the parameters that determine $h(t)$, especially the positional parameters $\Theta$ and $\Phi$, may be determined from the data. The method we will use is that of linear least-squares parameter estimation. In this method, we assume that approximate initial values have been found for the $m$ parameters $q_a$ that determine the signals. We also assume that $n$ observations $h_i$ of the waveform have been made and that the difference between the observed waveform and the waveform predicted by the current approximate values of the parameters is

$$y_i = h_{i,\text{observed}} - h_{i,\text{computed}} = \frac{\partial h_i}{\partial q_a}\Delta q_a + \nu_i \quad (3.22)$$

where $\nu_i$ is the noise associated with each measurement and $\Delta q_a$ is the error in the current value of the parameter $q_a$. The goal of least squares parameter estimation is to find the adjustments $\Delta q_a$ to the parameters that will minimize the $\nu_i$ errors in a least-squares sense. If we define

$$x_a \equiv \Delta q_a \text{ and } G_{ia} \equiv \frac{\partial h_i}{\partial q_a} \quad (3.23)$$

then the total squared error (also called the variance) may be written

$$\begin{aligned} V \equiv \nu_i\nu_i &= (y_i - G_{ia}x_a)(y_i - G_{ic}x_c) \\ &= y_iy_i - 2y_iG_{ia}x_a + G_{ia}G_{ic}x_ax_c \end{aligned} \quad (3.24)$$

where sums over $i$ from 1 to $n$ and over $a$ and $c$ from 1 to $m$ are implied by the repetition of the subscripts. The least squares requirement of minimizing the variance proceeds by requiring that

$$\frac{dV}{dx_b} = -2y_iG_{ib} + 2G_{ib}G_{ia}x_a = 0 \quad (3.25)$$

If we define an $m \times m$ square information matrix

$$A_{ab} = G_{ia}G_{ib} \quad (3.26)$$

then the information equation 3.25 may be written

$$A_{ab}x_a \equiv y_iG_{ib} \quad (3.27)$$

giving a solution for $x_a$ of

$$x_a = A_{ab}^{-1}y_iG_{ib} \quad (3.28)$$

where $A_{ab}^{-1}$ is the matrix inverse of $A_{ab}$.

The uncertainty in $x_a$ that results from such a procedure is related to the uncertainty $\sigma_y$ in the observations $y_i$. It is found by defining the covariance matrix as the expectation value of the mean-squared deviation of $x_a$ from its expected value:

$$C_{ab} \equiv \langle [x_a - \langle x_a \rangle][x_b - \langle x_b \rangle]\rangle \quad (3.29)$$

where $\langle ... \rangle$ represents the expectation value of the quantity inside the brackets. Since only $x_a$ and $y_i$ are random variables, the expectation value of $x_a$ is given by

$$\langle x_a \rangle = A_{ac}^{-1}G_{ic}\langle y_i \rangle \quad (3.30)$$

Combining equations 3.28, 3.30 and 3.29, we get

$$C_{ab} \equiv A_{ac}^{-1}G_{ic}A_{bd}^{-1}G_{jd}\langle[y_i - \langle y_i\rangle][y_j - \langle y_j\rangle]\rangle \quad (3.31)$$

If the errors in the observations are uniformly distributed and uncorrelated (see subsection III F), then the expectation value of the mean squared error in the observations is

$$\langle[y_i - \langle y_i\rangle][y_j - \langle y_j\rangle]\rangle = \delta_{ij}\sigma_y^2 \quad (3.32)$$

where $\delta_{ij}$ is the Kronecker delta. Substituting equation 3.32 into equation 3.31, and remembering that $A_{cd}A_{bd}^{-1} = \delta_{cb}$, we can write

$$C_{ab} = \sigma_y^2 A_{ab}^{-1} \quad (3.33)$$

According to equation 3.29 the expected errors $\sigma_a$ in the parameter values are given by the the diagonal elements of the covariance matrix $C_{ab}$:

$$\sigma_a = \sqrt{\langle[x_a - \langle x_a\rangle][x_a - \langle x_a\rangle]\rangle} \quad (3.34)$$

Therefore, equation 3.33 implies that the expected errors are equal to

$$\sigma_a = \sigma_y\sqrt{A_{aa}^{-1}} \quad (3.35)$$

(no sum over the index $a$ is implied here).

### E. Calculating the Derivatives

In order to construct the information matrix for the uncertainty calculation, we need to calculate the partial derivatives of the hypothetical observed signal $h_k(t)$ with respect to each of the nine unknown parameters $\tau, t_c, R_L, \delta, i, \phi_0, \psi, \Theta$, and $\Phi$. We evaluated all of these derivatives analytically to get specific expressions for $\partial h_k/\partial q$ (where $q$ is any one of the nine variables) as a function of time. This is mostly a matter of simple (though tedious) differential calculus. The following notes give an overview of the calculations.

Consider the derivative with respect to an arbitrary variable $q$. According to equation 3.11,

$$\begin{aligned}\frac{\partial h_k}{\partial q} = &\frac{\partial F_{+,k}}{\partial q}h_+ + F_{+,k}\frac{\partial h_+}{\partial q} \\ &+\frac{\partial F_{\times,k}}{\partial q}h_\times + F_{\times,k}\frac{\partial h_\times}{\partial q}\end{aligned} \quad (3.36)$$

Of the partials of the beam-pattern functions $F$, only $\partial F/d\Theta$, $\partial F/d\Phi$, and $\partial F/d\psi$ are nonzero, so evaluating these partials is relatively straightforward. If we look at equation 3.1, we see that

$$\frac{\partial h_{+,\times}}{\partial q} = \frac{\partial h_0}{\partial q}\left[\varepsilon^2 H_{+,\times}^{(0)} + \varepsilon^3 H_{+,\times}^{(1/2)} + \ldots + \varepsilon^6 H_{+,\times}^{(4/2)}\right]$$
$$+ h_0\left[2\varepsilon H_{+,\times}^{(0)} + 3\varepsilon^2 H_{+,\times}^{(1/2)} + \ldots + 6\varepsilon^5 H_{+,\times}^{(4/2)}\right]\frac{\partial \varepsilon}{\partial q}$$
$$+ h_0\left[\varepsilon^2 \frac{\partial H_{+,\times}^{(0)}}{\partial q} + \ldots + \varepsilon^6 \frac{\partial H_{+,\times}^{(4/2)}}{\partial q}\right] \quad (3.37)$$

where $h_0 \equiv 2\tau c\eta/5r$.

Now, many of the quantities appearing in this equation are written in terms of the variable $\eta$, but the more fundamental variable describing the difference between the masses involved in the binary is $\delta$. Since $\eta$ is a simple function of the fundamental variable $\delta$ alone, we can use the chain rule to evaluate partial derivatives with respect to $\delta$ as follows:

$$\frac{\partial f(\eta,\ldots)}{\partial \delta} = \frac{\partial f(\eta,\ldots)}{\partial \eta}\frac{d\eta}{d\delta} = \frac{\partial f(\eta,\ldots)}{\partial \eta}\left(-\frac{\delta}{2}\right) \quad (3.38)$$

Similarly, note that $h_0$ is stated in terms of the geometric distance to the source $r$, but we are taking the fundamental distance variable to be the luminosity distance $R_L$, so we evaluate the partial derivative of $h_0$ with respect to $R_L$ as follows:

$$\frac{\partial h_0}{\partial R_L} = \frac{\partial h_0}{\partial r}\frac{\partial r}{\partial R_L} = \frac{\partial h_0}{\partial r}\frac{1}{1+z} \quad (3.39)$$

The expansion parameter $\varepsilon$ explicitly depends on $t$, and (through the angular frequency $\omega$) on the variables $t_c$ and $\eta$ (or $\delta$), so the only partials we need to evaluate the second line of equation 3.37 are:

$$\frac{\partial \varepsilon}{\partial \tau} = \frac{\varepsilon}{3\tau} + \frac{\varepsilon}{3\omega_s}\frac{\partial \omega_s}{\partial \tau} \quad (3.40a)$$

$$\frac{\partial \varepsilon}{\partial \delta} = \frac{\varepsilon}{3\omega_s}\frac{\partial \omega_s}{\partial \delta} \quad (3.40b)$$

$$\frac{\partial \varepsilon}{\partial t_c} = \frac{\varepsilon}{3\omega_s}\frac{\partial \omega_s}{\partial t_c} \quad (3.40c)$$

The greatest challenge in equation 3.37 is evaluating the partials of the $H$s. These functions, which in the higher-order cases can become quite complex, depend only on $i$ and $\delta$ and/or $\eta$ explicitly, but also depend on the received phase $\phi_r$, which in turn depends on $\tau, t_c, \delta, \phi_0, \Theta$ and $\Phi$. Therefore, generally we have to evaluate, for each of the ten $H$ terms,

$$\frac{\partial H}{\partial i}, \frac{\partial H}{\partial \delta}, \text{ and } \frac{\partial H}{\partial q} = \frac{\partial H}{\partial \phi_r}\frac{\partial \phi_r}{\partial q} \quad (3.41)$$

for $q = \tau, t_c, \delta, \phi_0, \Theta,$ and $\Phi$.

Evaluating the derivatives of $\phi_r$ is complicated by the fact that one cannot easily evaluate the Doppler-shift integral in equations 3.8 analytically because of the complicated time-dependence of the source-frame orbital frequency $\omega_s$. Therefore, we expressed each of the partial derivatives of $\phi_r$ in terms of an integral equation: for example

$$\frac{\partial \phi_r}{\partial \tau} = \frac{1}{1+z}\left[\frac{\partial \phi}{\partial \tau} - \frac{\Omega R}{c}\sin\Theta \int_0^t \frac{\partial \omega_s}{\partial \tau}\sin(\Omega t - \Phi)\,dt\right] \quad (3.42a)$$

$$\frac{\partial \phi_r}{\partial \Theta} = \frac{1}{1+z}\left[-\frac{\Omega R}{c}\cos\Theta \int_0^t \omega_s \sin(\Omega t - \Phi)\,dt\right] \quad (3.42b)$$

We then evaluated these integrals at each time step using a fourth-order Runge-Kutta integration scheme. Since $\omega_s$ is a fairly slowly-varying function of time (except in the final stages of coalescence, where the whole post-Newtonian power-series approximation breaks down anyway), the Runge-Kutta approach should yield excellent estimates for the values of these integrals.

### F. Noise Curves

Ultimately, the detector's position sensitivity is determined by the $\sigma_y$ noise in the detector. We took the total noise for the covariance study to be

$$\sigma_y^2 = S_n(f)\,\Delta f \quad (3.43)$$

where the bandwidth $\Delta f = 1/2dt$ ($dt$ being the time between samples), and $S_n(f)$ is the spectral noise density. The total rms noise in the detectors, sampled at intervals $dt$, would of course be the integral of $S_n(f)$ over the bandwidth. By taking instead the simple product of $\Delta f$ and $S_n(f)$ at the primary signal frequency, we account for the temporal correlations created in the sampled data by the larger low-frequency noise components (see Appendix for details).

To estimate the spectral noise density $S_n(f)$, we have taken the published noise curves for the proposed space missions. For the ecliptic-plane scenario, we used the noise curve for OMEGA; for the precessing-plane case, we use the noise curve for LISA. The instrumental noise curves for both missons are dominated by two sorts of noise, acceleration noise and position noise. The generation of sensitivity curves with these types of noise present is discussed in Larson, Hiscock, and Hellings [5]. In addition, all instruments in the low-frequency band will have to contend with a background of clutter formed by unresolvable close compact binaries in the Galaxy [6]. The total spectral noise density function is thus given by

$$S_n(f) = [S_a(f)f^{-4} + S_x(f)](1 + 2\pi fT)^2 + S_c(f) \quad (3.44)$$

where $S_a(f)$ is the strain noise produced by acceleration noise, $S_x(f)$ is the strain noise produced by position noise, and $T$ is the light-travel time between detectors. The $(1 + 2\pi fT)^2$ factor arises because the sensitivity of the instrument to gravitational waves is reduced when the gravitational wave wavelength is less than the baseline distance between spacecraft. The values of the terms in equation 3.44 are given in Table I. In addition, in both missions we are assuming that we are looking at clutter noise of

$$S_c(f) = \begin{cases} (2.07 \times 10^{-43} \text{ Hz}^{0.9})f^{-1.9} & f < f_1 \\ (4.73 \times 10^{-61} \text{ Hz}^{6.5})f^{-7.5} & f_1 < f < f_2 \\ (1.41 \times 10^{-47} \text{ Hz}^{1.6})f^{-2.6} & f > f_2 \end{cases} \quad (3.45)$$

where $f_1 = 7.1 \times 10^{-4}$ Hz and $f_2 = 1.8 \times 10^{-3}$ Hz. The total spectral noise curves resulting from equations 3.44 and 3.45 are shown in Figure 2.

### G. Summary of the Algorithm

To perform the actual calculations, we wrote a FORTRAN program that does the following:

1. The program reads values for the fundamental parameters $m_1 + m_2, t_c, R_L$ (or $z$), $\delta, i, \phi_0, \psi, \Theta$, and $\Phi$ for a hypothetical source from an input file. In addition, the input file also specifies the size of the time step $dt$ to be used and the detector array's initial orientation angle $\alpha_0$. (We generally used $dt = 60$ s in our runs so that $dt$ was somewhat smaller than the shortest period for waves that either detector can easily register.)

2. The program then uses the specified values to set up a number of constants used in the calculation, and initializes the Runge-Kutta scheme used to evaluate the integrals in the expressions for $\phi_r$ and its partial derivatives.

3. It then executes a loop that at each time step (including $t = 0$)

   - calculates the noise level $S_n$ at the current observed gravity wave frequency,
   - calculates values of $\phi_r$ and its partial derivatives,
   - calculates $h_k(t)$ and its partial derivatives for each detector,
   - adds the information from the current time step to matrix $A^*_{ab} = A_{ab}/\sigma_y^2$ (where $A_{ab}$ is as defined in equation 3.26) as follows:

   $$A^*_{ab}(t) = A^*_{ab}(t - dt) + \frac{dt}{S_n(f)} \sum_{k=1,2} \frac{\partial h_k(t)}{\partial q_a} \frac{\partial h_k(t)}{\partial q_b} \quad (3.46)$$

   where $q_i$ corresponds to the $i$-th fundamental parameter.
   - evaluates the Doppler-shift integrals needed for the next time step.

   The loop ends when the time to coalescence is smaller than 10 times the current orbital period or after one full year of observation, whichever occurs sooner.

4. The program then uses a standard matrix-inversion package to invert the matrix $A^*_{ab}$ to get $C_{ab} = \sigma_y^2 A_{ab}^{-1} = (A^*_{ab})^{-1}$ (see equation 3.33). According to equation 3.35, the diagonal elements of this inverted matrix correspond to the uncertainties in the values of the fundamental parameters that we could determine from the wave train generated by the hypothetical source in question. If the matrix cannot be inverted, the program eliminates the parameter causing the problem and computes uncertainties for the remaining parameters. Where possible, the program simultaneously estimates uncertainties for *all nine* of the fundamental parameters, not just the angular position parameters.

5. It then displays the calculated results for the uncertainties, including the solid-angle uncertainty in the angular position of the source.

Running this program repeatedly with different parameter values allows us to learn about how the parameter uncertainties might depend on the characteristics of the source.

We tested the program by making certain that in the low-mass limit, its results agreed completely with a much simpler program we had written earlier to handle the case of strictly monochromatic waves from low-mass binaries. Even though significant differences between our assumptions and those of Cutler [1] and Cutler and Vecchio [2,3] made exact comparisons difficult, we also tested our program by comparing its results to their published results for the precessing-plane case, and found agreement typically within a factor of three.

### IV. RESULTS

The results in principle depend on ten parameters: the nine fundamental (unknown) parameters and the presumably known initial orientation parameter $\alpha_0$. Obviously, it is difficult to sample such a large parameter space and display its results meaningfully. We did find that in spot checks the uncertainty in the source's angular position seemed to be generally insensitive to the parameters $\phi_0$, $\psi$, $\alpha_0$, and $\Phi$, and that the results were basically the same for polar angles $\Theta$ above and below the ecliptic that made the same angle with the ecliptic. In this paper, we will focus primarily how the angular position uncertainty depends on the total mass of the source, or its frequency, and the polar angle $\Theta$.

Figure 3 shows the base-ten logarithm of the angular position uncertainty (expressed in steradians) as a function of $\Theta$ for essentially monochromatic compact binaries at selected frequencies ranging from 0.1 mHz to 100 mHz. The masses were chosen so that $t_c$ was long compared to a year, the distance was chosen so that the signal-to-noise ratio in the ecliptic case was roughly 10, and the inclination was chosen so that $\cos i$ was about 0.8.

Note that in the ecliptic-plane case (solid curves), there is a monotonic improvement in angular position resolution with increasing frequency, as the Doppler shift becomes easier to track. Indeed, in the strictly monochromatic limit, the angular position uncertainty should fall strictly as $f^{-2}$ (see page 99 of [2]), and our results are consistent with this. Note also that the angular position uncertainty increases as one approaches $\Theta = 90°$ (that is, the plane of the ecliptic). This is because in the ecliptic plane case, the only source of information for $\Theta$ comes via the Doppler shift (see equation 3.8a). Since this effect is proportional to $\sin \Theta$ the partial derivative with respect to $\Theta$ is proportional to $\cos \Theta$ which goes to zero at $\Theta = 90°$. There is thus no $\Theta$ information in the information matrix at $\Theta = 90°$.

In the precessing-plane case (the dotted curves), there is moderately good angular resolution even at low frequency because of the extra information provided by the precession of the array plane. The angular resolution improves as frequency increases because of the additional information provided by the Doppler shift. The angular resolution in this case proves to be relatively independent of $\Theta$ (the precession of the array plane means that the ecliptic plane is not as special as it is in the ecliptic case).

Figure 4 shows the angular uncertainty as a function of $\Theta$ for various large-mass coalescing sources. All of these sources have equal-mass partners (at $10^7$, $10^6$, $10^5$, and $10^4$ solar masses respectively), are located at redshift $z = 1$, have an inclination whose cosine is 0.8, and are observed for the last year before coalescence. Note that in the precessing-plane case (the dotted curves) the angular resolution is fairly good even for the lowest-frequency cases (the cases with the largest masses). The resolution stays relatively constant as the mass decreases, because even as the signal strength drops, the noise curve also drops. Again the angular position uncertainty is relatively independent of $\Theta$.

In the ecliptic-plane case, the angular resolution again improves as the frequency increases (as expected), but curiously, and in sharp contrast to the monochromatic case, the angular resolution improves significantly as $\Theta$ approaches $90°$. We will discuss this further in the next section.

Figure 5 shows graphs for middle-mass black holes falling into supermassive black holes. One can see that the results are qualitatively similar to the equal-mass black-hole mergers shown in Figure 4.

## V. DISCUSSION OF THE RESULTS

In the ecliptic plane case, the angular position uncertainty decreases significantly as we approach the ecliptic plane in the massive black hole cases shown in Figure 4, but the uncertainty increases significantly in the monochromatic case shown in Figure 3. What could be so different about the two cases? It turns out that the crucial difference in determining this effect is that in the massive black hole case, the higher-order post-Newtonian corrections to the waveform amplitude are significant (because the expansion parameter $\varepsilon$ in the power series is large), while in the monochromatic case, they are not. Figure 6 clearly shows that when we artificially turn off the higher-order terms, the angular position uncertainty increases with $\Theta$ for the massive black-hole case, just as it does for the monochromatic case. Our runs show that virtually all of the difference comes from the lowest-order nonzero harmonic above the fundamental: adding *only* this term to the fundamental produces essentially the same curve we get when we include *all* higher-order terms.

Why is this? We may consider the waveforms seen at each of the two detectors as depending on three effects. The first effect is the monotonic increase in the frequency of the source, as given by equations 3.9. If this increase is written by expanding the frequency in a Taylor series, then the behavior can be expressed in terms of the derivatives $\omega_0, \dot{\omega}_0, \ddot{\omega}_0$, etc. As seen in equations 3.9, these derivatives are linked to the basic variables $\eta$, $\tau$, and $t_c$. Observation of the time series will thus determine the frequency derivatives and thus $\eta$, $\tau$, and $t_c$ independently of any other features of the waveforms. The second effect is the variation of the waveform with orientation of the detector, as given by equations 3.15. These form factors depend explicitly on $\Theta$, $\Phi$ (through $\alpha_k$), and $\psi$. The form factors may not be seen directly in the waveform, but only in convolution with the third effect, the amplitudes of the two polarizations of the waves, as given by equation 3.1. These $h$-functions are determined by the already-known $\tau$, $\eta$, and $\omega$, and by the unknown parameters $r$, $i$, and $\phi_0$. There are thus six unknown parameters that must be determined from the time series alone, without any help from the change in frequency with time. If only the fundamental frequency of the gravitational wave were present (equations 3.3a and 3.3b), then each detector would see only a single gravitational wave frequency whose amplitude and phase would be the only observables. For two detectors, there would be two amplitude observables and two phase observables, but this would not be enough to determine the six unknown parameters, $\Theta$, $\Phi$, $\psi$, $r$, $i$, and $\phi_0$. However, if a second harmonic of the wave is included (the harmonics $H^{(1/2)}$ of equations 3.3c and 3.3d), then a Fourier analysis of the detected signals would determine amplitudes and phases for both harmonics. Since the mix of phases and amplitudes between the two harmonics depends on $i$ and $\phi_0$, there will be nontrivial information in these additional amplitudes and phases, and all six gravitational wave parameters can be determined from these eight signal response parameters. The ability of the higher harmonics to add information, however, is quenched as $\Theta \to 0$, because the form factors depend on $\Theta$ only through $\cos\Theta$ and also because $\Phi$ and $\psi$ become degenerate near the pole of the ecliptic.

In any case, Figure 6 makes it clear that including the higher-order terms makes a huge difference for values of $\Theta$ close to the ecliptic plane. It is worth noting in this context that if massive-black hole sources are distributed evenly about the sky, half will lie within $\pm 30°$ of the ecliptic plane, where the effect of including the higher-order terms is most significant.

This result has not been previously reported in the literature. Cutler [1] and Cutler and Vecchio [3] only model the gravitational waveform itself to lowest post-Newtonian order, leading them to conclude (appropriately, considering the limitations of that model) that a detector array in ecliptic-plane configuration "has essentially no angular and distance resolution" for massive black hole mergers (see page 105 of [3]). However, the results reported here indicate that we need not be so pessimistic: indeed, for $\Theta$ very near the ecliptic, the angular resolutions in the ecliptic-plane and precessing-plane cases are quite comparable as long as we include the higher-order terms.

It is interesting to note that we do not observe anything comparable for the precessing-plane case: we find that artificially suppressing the higher-order terms in the waveform does not change the angular uncertainties very much. In the precessing plane case, the fact that the normal to the detector plane moves around the sky means that $\theta_D$ and $\psi_D$ depend on time in a characteristic way even though $\Theta$ and $\psi$ are fixed. This provides information that allows one to disentangle the effects of the latter parameters from those of $i$ and $R_L$ even without the extra information provided by the higher-order terms in the wavefunction.

## VI. CONCLUSIONS

While in the main the results from this study validate much the work that has already been done on the angular position resolution of space-based detectors, this work underlines the unexpected but critically important role that higher-order post-Newtonian corrections play in determining parameters in massive black-hole mergers, particularly for a detector array in the ecliptic-plane configuration. This suggests that future research into how one determines parameters of massive black-hole mergers from an observed wave train should definitely take account of these higher-order post-Newtonian corrections, at least if one is considering an ecliptic-plane detector configuration.

It is also worth noting that our results indicate that neither detector configuration is able to yield an angular position uncertainty much less than about $10^{-4}$ steradians for massive black-hole mergers (see Figures 4 and 5). For comparison, the full moon occupies a solid angle of about $6 \times 10^{-5}$ sr. This means that even though either kind of detector can easily register massive-black-hole mergers at cosmological distances, neither will be able to locate the source precisely enough (by several orders of magnitude) to indicate the galaxy where the merger is taking place. One can therefore only locate the merger if one can also observe the electromagnetic signature of such an event. To our knowledge, little is yet definitively known about what the electromagnetic signature of a merger might look like. Using the position information provided by either kind of detector in practice (or even knowing whether the information has any practical use at all) will thus require more research into the astrophysics of black-hole mergers.

## ACKNOWLEDGMENTS

We wish to thank the referee for finding an error in the original version of this paper, which has been corrected in this version.

## APPENDIX: HANDLING RED NOISE

The gravitational wave sources we have considered in this paper are all found at frequencies on the left of Figure 2, where the noise spectrum is falling roughly like $f^{-4}$. Noise such as this, where there is more power at low Fourier frequencies than at high frequencies, is termed "red noise". It is generally known that parameter estimation, including spectral estimation, produces biased estimates when the background noise is not white. [7,8] The solution to this bias problem lies in a process called "prewhitening". In this process, the data are passed through a filter that generates a new time series whose noise spectrum is flat, and the parameter estimation is performed using this filtered data. The estimates will then be bias-free. In this appendix, we show how to calculate the signal-to-noise ratio (SNR) for such a process, assuming the signal is dominated by a single frequency.

Let us consider a time series $y(t) = h(t) + n(t)$, where $h(t)$ is the signal and $n(t)$ is the red noise. To prewhiten the data, the time series is passed through a liner filter to produce:

$$x(t) = F(t) * y(t) \equiv \int_0^T F(t-\tau) y(\tau) d\tau \qquad (A1)$$

We write the effect of this filter on the signal $h(t)$ as $g(t) = F(t) * h(t)$ and on the noise as $m(t) = F(t) * n(t)$. The filter $F$ is chosen so that the power spectrum of $m(t)$ will be flat. Since the filter is a convolution integral, the effect of the filter in frequency space yields simple products, $g(f) = F(f)h(f)$ and $m(f) = F(f)n(f)$, where $F(f)$ is called the transfer function of the filter. [For example, if the power spectrum of the detector noise $S_n(f) = n(f)^2$ were exactly proportional to $f^{-4}$, then the linear filter would be $F(t) = d^2/dt^2$ whose transfer function is $F(f) = 4\pi^2 f^2$. The power spectrum of $m(t)$ would therefore be flat.]

The SNR for the filtered data is given by $(\text{SNR})^2 = \langle g^2(t)\rangle/\langle m^2(t)\rangle$, where the angular brackets denote a time average. The mean squared signal and noise strengths may be written in terms of their spectral densities to give

$$(\text{SNR})^2 = \frac{\int_{f_L}^{f_H} S_g(f)df}{\int_{f_L}^{f_H} S_m(f)df}$$
$$= \frac{\int_{f_L}^{f_H} F^2(f)S_h(f)df}{S_m(f_H - f_L)} \quad (A2)$$

where, in the last step, the fact that $S_m(f) = \text{const}$ has been used to complete the integral in the denominator and the fact that $g(f) = F(f) * h(f)$ has been used to expand the numerator.

Now let us assume that the signal $h(t)$ is essentially monochromatic, at frequency $f_0$, so that in the numerator we will have $S_h(f) = \delta(f-f_0)S_h(f_0)$. We also assume that the data have been previously high-pass filtered with cutoff at $f_L = f_0$. In the denominator, because $S_m(f)$ is constant, its relation to $S_n(f)$ may be worked out at any frequency desired. If we choose $f_0$ as that frequency, then we have

$$(\text{SNR})^2 = \frac{F^2(f_0)\int_{f_0}^{f_H} \delta(f-f_0)S_h(f_0)df}{F^2(f_0)S_n(f_0)(f_H - f_0)}$$
$$= \frac{\langle h(t)^2\rangle}{S_n(f_0)(1/2\Delta t)} \quad (A3)$$

where, in the last step, we have recognized $\langle h(t)^2\rangle$ as the integral of the spectral density $S_h$ and $f_H = 1/2\Delta t$ as the Nyquist frequency, assumed to be much higher than $f_0$. This is the formula we use for noise in our detector, as given in equation 3.43.

FIG. 1. Spacecraft configurations for the ecliptic-plane and precessing-plane options. Figure 1a shows top and side views of the three detector spacecraft in the ecliptic-plane option. The arrow shows the normal vector to the plane of the orbit, which is fixed in this case. Figure 1b shows top and side views of the three detector spacecraft in the precessing-plane option. In this case, the spacecraft do not orbit the earth but rather orbit the sun in such a way that the satellites form an apparently rigid equilateral triangle that rotates clockwise around the array's center, which in turn orbits the sun. The plane of the array is tilted 60° from the plane of the ecliptic (which is why the circular trajectories of the satellites look elliptical in the top view). The projection on the ecliptic of the array's normal vector always points toward the sun.

FIG. 2. Graph of the noise spectral density $S_n(f)$ that was used in the covarance analysis. The solid curve is for the ecliptic-plane case (using OMEGA noise estimates) while the dashed curve is for the precessing-plane case (using LISA noise estimates).

FIG. 3. Angular uncertainty as a function of $\Theta$ for a compact binary emitting gravitational waves with various different (but essentially constant) frequencies. The stellar masses were chosen in each case so that $t_c$ was much longer than one year, and the distance was chosen so that the signal-to-noise ratio was about 10 in the ecliptic-plane case (the signal amplitudes are specified in each picture). The other parameters were set somewhat arbitrarily as follows: $\alpha_0 = 0$, $i = 36.9°$ ($\cos i = 0.8$), $\Phi = 268.5° = 4.69$ rad, $\psi = 114.6° = 2.0$ rad.

FIG. 4. Angular uncertainty as a function of $\Theta$ for various equal-mass pairs of supermassive black holes undergoing coalescence. In each case, the coalescing pair was assumed to be at redshift $z = 1$ and the time to coalescence $t_c$ was chosen to be one year. The initial gravitational wave frequency shown in each graph is the frequency at the beginning of the one-year observational period, and the final frequency is the frequency when the calculation ended roughly 10 orbits before coalescence. The other parameters were set as in Figure 3.

FIG. 5. Angular uncertainty as a function of $\Theta$ for two situations where a medium-mass black hole coalesces with a supermassive black hole. Other parameters were set as specified in Figure 4.

FIG. 6. Angular uncertainty as a function of $\Theta$ for coalescing massive black holes, each with a mass of $10^6$ solar masses (this is the same situation shown in Figure 4b). This illustrates that the higher-order terms in the post-newtonian approximation for the gravitational waveform have a large effect on the angular resolution in the ecliptic-plane for values of $\Theta$ near the ecliptic plane: artificially suppressing these terms leads to very poor resolution near the ecliptic plane, while including them leads to an angular resolution that is nearly the same as for the precessing-plane case when $\Theta$ is close to the ecliptic. It turns out that virtually all of the difference is due to the lowest-order nonzero harmonic term above the fundamental: adding this term alone produces a curve essentially identical to the lower curve shown here.

TABLE I. Table showing the magnitudes of the spectral noise constants in equation 3.44 for the OMEGA and current LISA detector configurations.

| Term | OMEGA | LISA |
|---|---|---|
| $S_a(f)$ | $5.8 \times 10^{-51}$ Hz$^3$ | $2.3 \times 10^{-52}$ Hz$^3$ |
| $S_x(f)$ | $1.26 \times 10^{-41}$ Hz$^{-1}$ | $1.26 \times 10^{-41}$ Hz$^{-1}$ |
| $2\pi T$ | 21 s | 100 s |

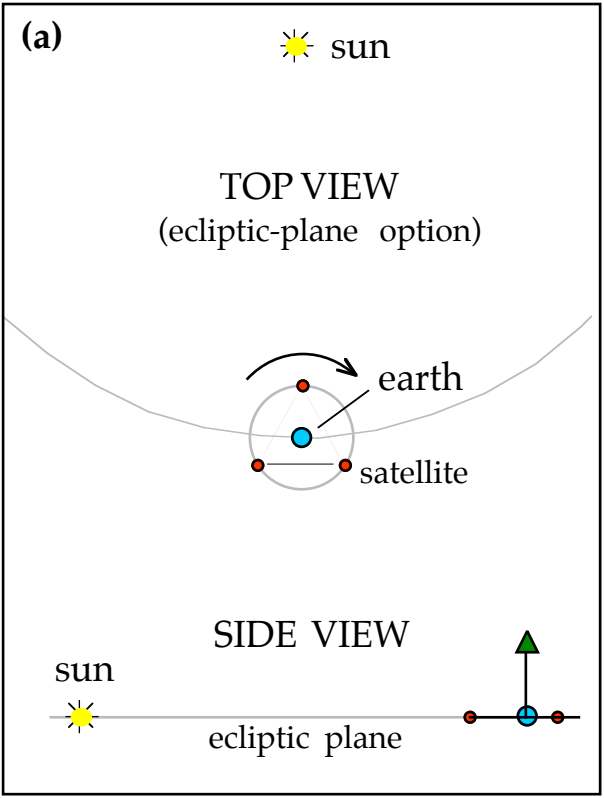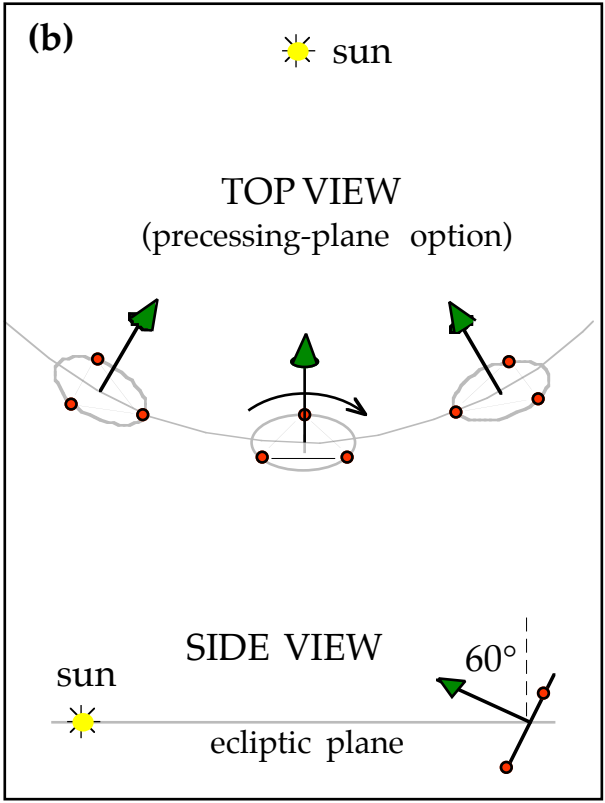

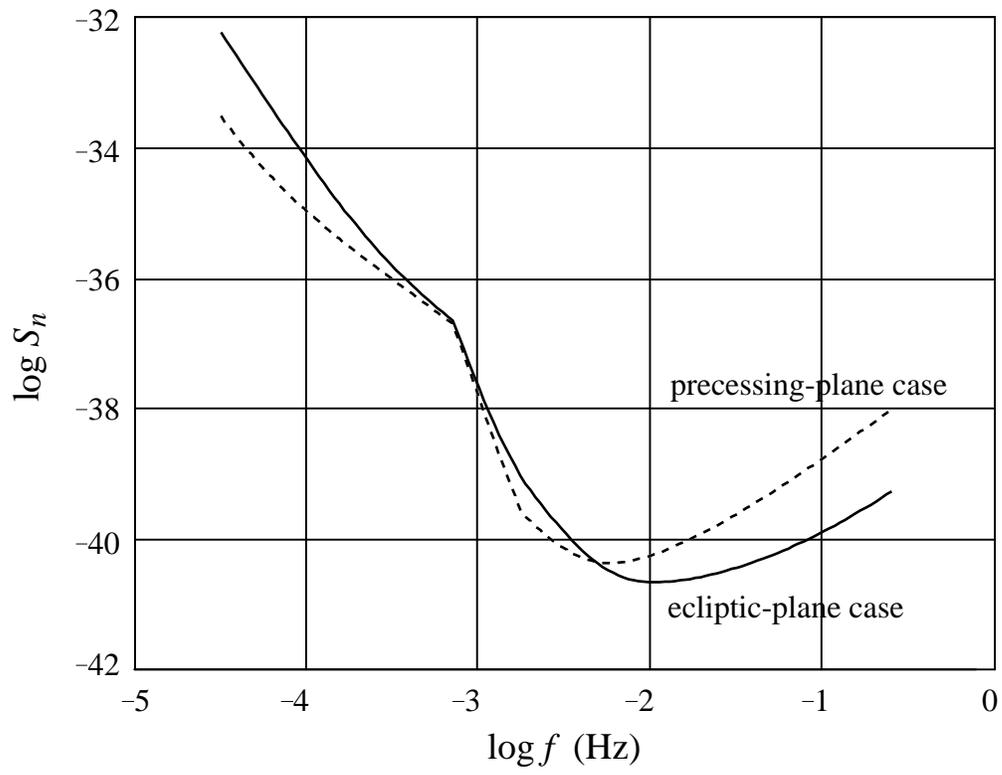

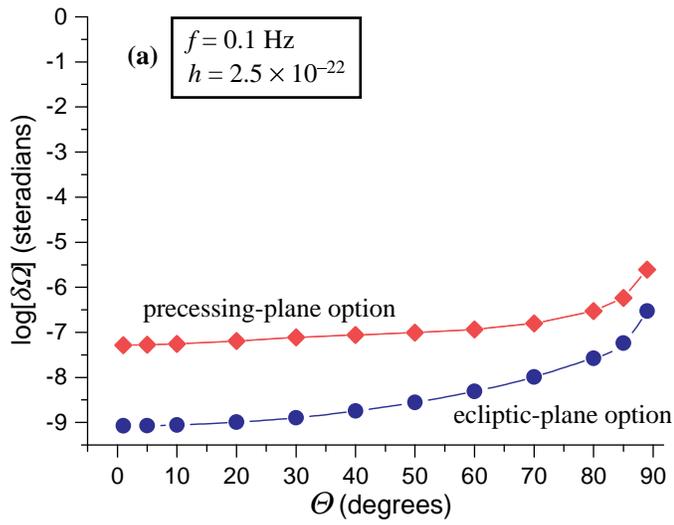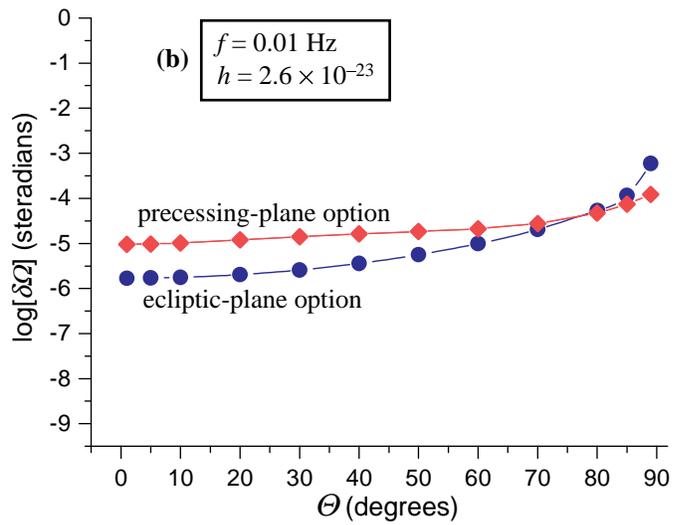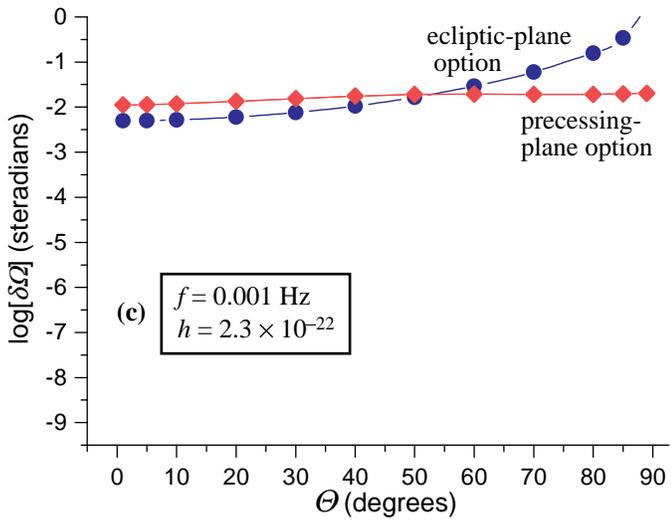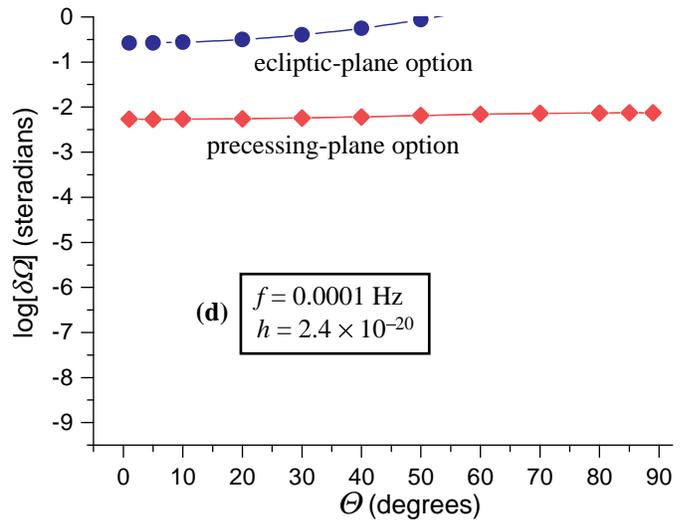

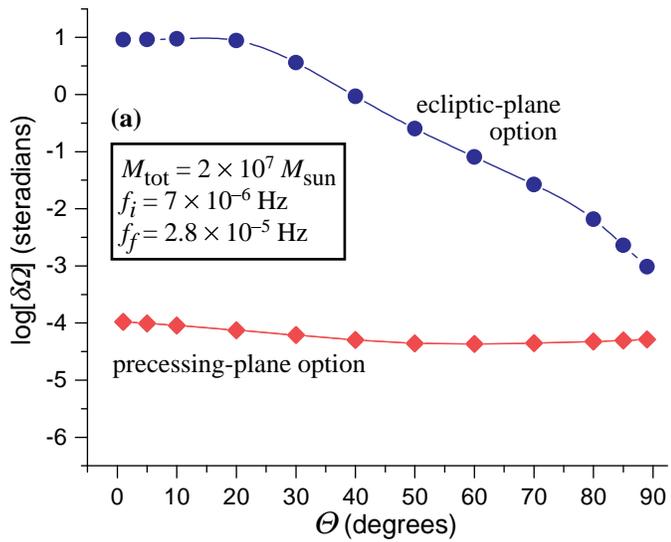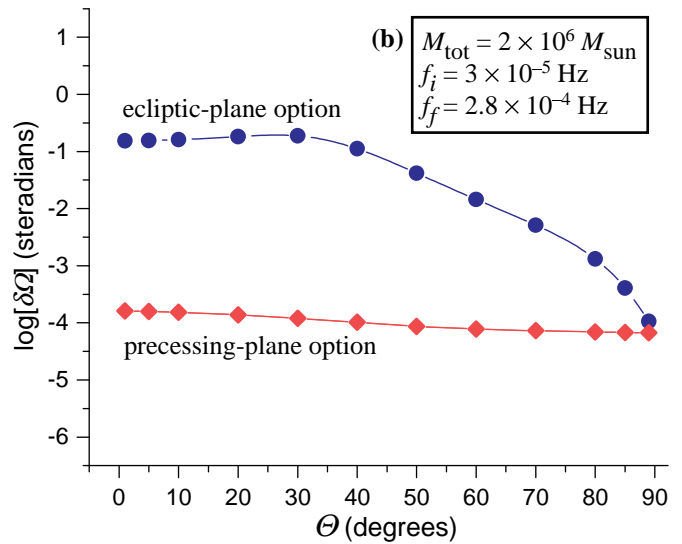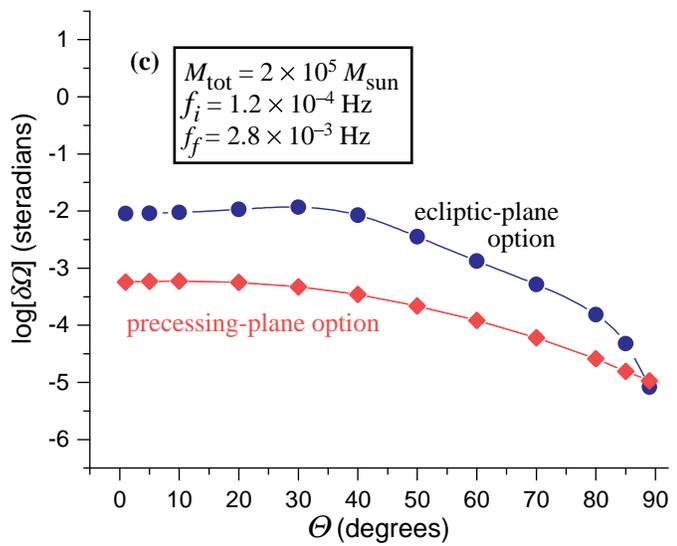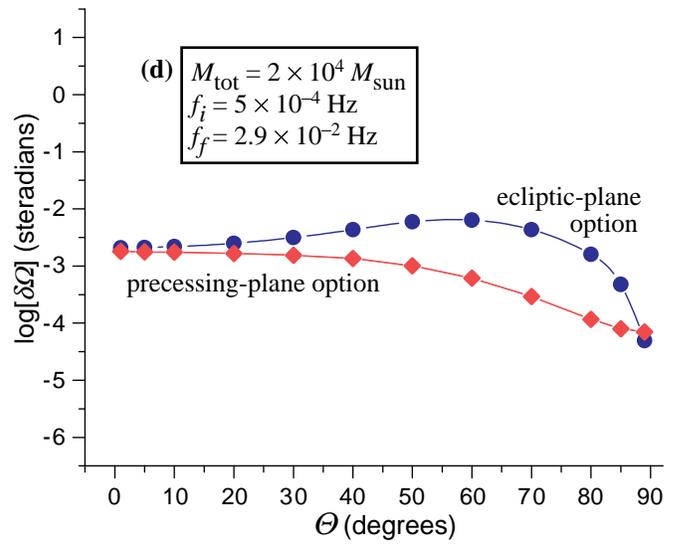

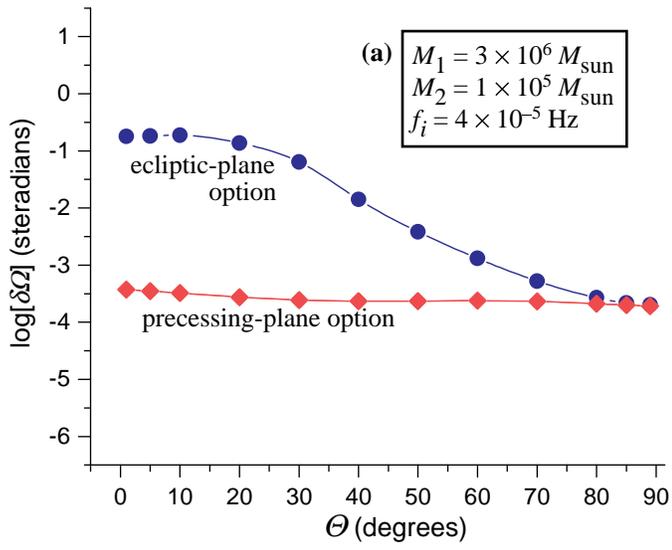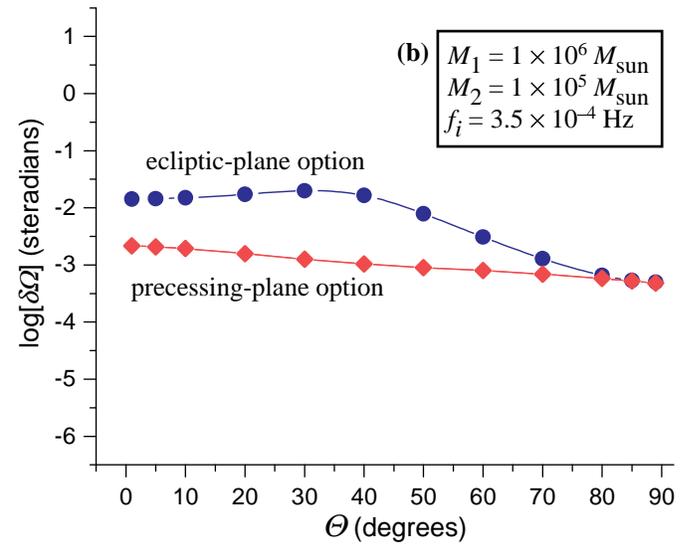

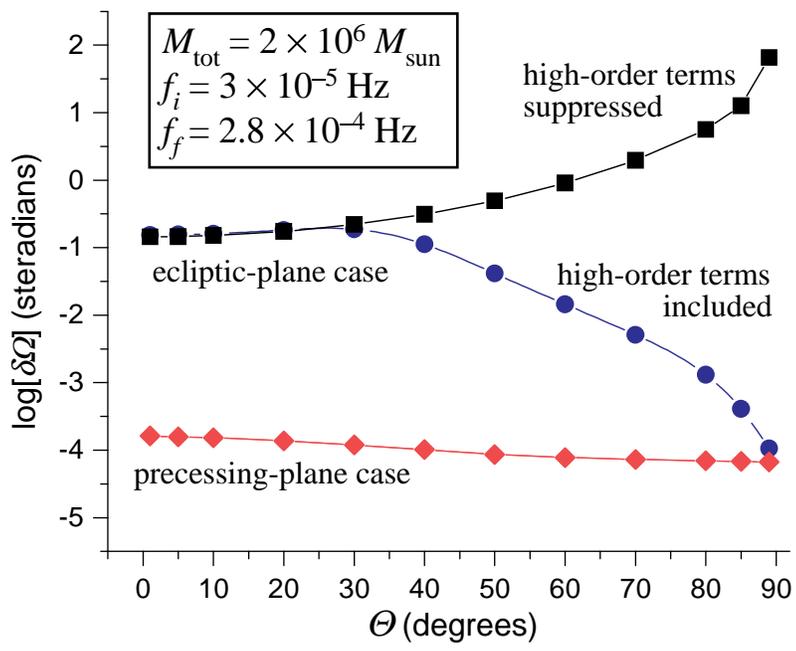